\documentclass[aps,pre,twocolumn,citeautoscript,superscriptaddress,byrevtex,nofootinbib,nobalancelastpage,floatfix]{revtex4}
\usepackage[pdftex]{graphicx}         
\usepackage{fancyhdr}                 
\usepackage{amsmath}                  
\usepackage{bm}                       
\usepackage{textcomp}
\usepackage[colorlinks=true,
    citecolor=blue,
    linkcolor=blue,
    urlcolor=blue,
    pagebackref=false]{hyperref}
\usepackage{bm}         
\usepackage{setspace}
\usepackage[thickspace,amssymb,detect-all]{SIunits}
\usepackage{multirow}
\usepackage[compact]{titlesec}
\usepackage{palatino}  
\usepackage{cases}

\begin{document}

\title{Cryogenic positioning and alignment with micrometer precision in a magnetic resonance force microscope}

\author{Corinne E. Isaac}
\affiliation{Department of Chemistry and Chemical Biology, Cornell University, Ithaca, New York 14853-1301, USA.}

\author{Elizabeth A. Curley}
\affiliation{Department of Chemistry and Chemical Biology, Cornell University, Ithaca, New York 14853-1301, USA.}

\author{Pam{\'e}la T. Nasr}
\affiliation{Department of Chemistry and Chemical Biology, Cornell University, Ithaca, New York 14853-1301, USA.}

\author{Hoang L. Nguyen}
\affiliation{Department of Chemistry and Chemical Biology, Cornell University,
Ithaca, New York 14853-1301, USA.}

\author{John A. Marohn}
\email[Email:]{jam99@cornell.edu}
\affiliation{Department of Chemistry and Chemical Biology, Cornell University, Ithaca, New York 14853-1301, USA.}

\begin{abstract}
    Aligning a microcantilever to an area of interest on a sample is a critical step in many scanning probe microscopy experiments, particularly those carried out on devices and rare, precious samples.
    We report a series of protocols that rapidly and reproducibly align a high-compliance microcantilever to a $<$10 $\micro\meter$ sample feature under high vacuum and at cryogenic temperatures.
    The first set of protocols, applicable to a cantilever oscillating parallel to the sample surface, involve monitoring the cantilever resonance frequency while laterally scanning the tip to map the sample substrate through electrostatic interactions of the substrate with the cantilever.
    We demonstrate that when operating a cantilever a few micrometers from the sample surface, large shifts in the cantilever resonance frequency are present near the edges of a voltage-biased sample electrode.
    Surprisingly, these ``edge-finder'' frequency shifts are retained when the electrode is coated with a polymer film and a $\sim$10 $\nano\meter$ thick metallic ground plane.
    The second series of methods, applicable to any scanning probe microscopy experiment, integrate a single-optical fiber to image line scans of the sample surface.
    The microscope modifications required for these methods are straightforward to implement, provide reliable micrometer-scale positioning, and decrease the experimental setup time from days to hours in a vacuum, cryogenic magnetic resonance force microscope.
\end{abstract}
\date{\today}
\maketitle

\section{Introduction}

The development of high-compliance microcantilevers \cite{Stowe1997jul,Jenkins2004may} and nanowires \cite{Nichol2008nov,Nichol2012feb} capable of observing attonewton-scale forces has enabled a number of interesting studies.
These include the observation of near-surface electric-field fluctuations over metals \cite{Stipe2001aug} and dielectrics \cite{Kuehn2006apr,Yazdanian2009jun},
the detection of electron-spin resonance from a single buried spin in quartz \cite{Rugar2004jul},
nuclear magnetic resonance imaging of a polymer film \cite{Nichol2013sep,Rose2017jul} and an individual virus \cite{Degen2009feb} with 2 and 4 to $10\;\nano\meter$ resolution, respectively, and the observation of spin-lattice relaxation in real time \cite{Vinante2011dec,Alexson2012jul,Isaac2016apr,Wagenaar2016jul}.
While each of these proof-of-concept experiments  represents a notable milestone, each experiment also required significant simplifications to enable its success.
The sample in the fluctuation studies \cite{Stipe2001aug,Kuehn2006apr,Yazdanian2009jun}, for example, was translationally invariant.
The virus sample in the imaging experiment of Ref.~\citenum{Degen2009feb} was affixed to the leading edge of a fragile microcantilever; it took weeks of scanning to locate the virus via its magnetic resonance signal.\footnote{Dan Rugar, personal communication.}
The creation of high-compliance microcantilevers with integrated nanomagnet tips \cite{Hickman2010nov,Longenecker2011may,Longenecker2012nov} creates exciting opportunities for measuring charge and spin fluctuations in patterned  thin-film devices \cite{Liu1998jun,Stowe1999nov,Cockins2012feb,Lekkala2012sep,Lekkala2013nov,Schumacher2016apr} and observing magnetic resonance in both devices and optically labeled, flash-frozen biological samples \cite{Orlova2011dec}.
To realize these opportunities in a magnetic resonance force microscope (MRFM), we must first position the cantilever over a chosen micrometer-scale feature on a centimeter-scale sample substrate.

To achieve high sensitivity cantilever detection of magnetic resonance via MRFM, experiments should be carried out in vacuum at a temperature near 4.2 kelvin \cite{Longenecker2012nov,Nichol2013sep,Isaac2016apr} or below \cite{Rugar2004jul,Degen2009feb,Vinante2011dec,Wagenaar2016jul}.
In order to detect a spin signal, transverse radiofrequency magnetic fields of millitesla strength or microwave frequency magnetic fields of microtesla strength must be applied to modulate nuclear or electron spins, respectively.
These transverse oscillating magnetic fields have traditionally been applied at a single frequency using a tuned coil \cite{Garner2004jun,Rugar2004jul,Mamin2007may} or with a microstripline half-wave resonator \cite{Moore2009dec,Hickman2010nov}.
However, at cryogenic temperatures, achieving high strength fields with these devices is limited as they must operate using only a few milliwatts of input power to avoid sample heating.
To address these stringent requirements, Poggio {\it {et al.}} \cite{Poggio2007jun}, demonstrated the detection of nuclear magnetic resonance with a 1 $\micro\meter$ wide microwire capable of generating $> 4 \; \milli\tesla$ of field at $115 \; \mega\hertz$ while dissipating just $350 \; \micro\watt$ of power.
These microwires have been used in a number of notable magnetic resonance experiments \cite{Degen2007dec,Longenecker2012nov,Vinante2011dec} including the nanometer-scale virus imaging experiment of Degen and coworkers \cite{Degen2009feb}.
To further increase the applications of these microwire devices, Isaac {\it{et al.}} demonstrated that by integrating a microwire into a 50-$\ohm$ coplanar waveguide, the devices can be operated at frequencies ranging from 1 $\hertz$ to 40 $\giga\hertz$, enabling the detection of nuclear magnetic resonance and electron spin resonance with a single source of transverse field \cite{Isaac2016apr}.
While these microwire and coplanar waveguide devices provide adequate transverse fields for a variety of magnetic resonance experiments on a limited power budget, they introduce an additional challenge --- aligning a high-compliance microcantilever to a 1 to 10 micrometer wide wire.
	
To avoid snap-in to contact, high-compliance cantilevers must be operated in the ``hangdown'' geometry \cite{Stowe1997jul,Nichol2012feb} --- oscillating parallel to the sample surface, with the cantilever's long axis oriented along the surface normal.
This geometry precludes using the cantilevers in conventional atomic-force-microscope mode to locate sample features.
Additionally, upon cooling to cryogenic temperatures any positioning strategy implemented must contend with misalignment of the cantilever and sample, which can exceed 100's of micrometers, due to thermal contraction of the microscope assembly.
Finally, to minimize sample-induced frequency noise \cite{Yazdanian2008jun,Yazdanian2009jun,Hoepker2011oct,Longenecker2012nov,Lekkala2013nov} it is often advantageous to coat the thin-film sample with an anti-static metal film \cite{Garner2004jun,Moore2009dec}; this coating creates the additional challenge of locating a sample feature buried beneath a metal ground plane.
	
Here we describe a series of protocols developed to rapidly align a scanned, high-compliance microcantilever operating in the ``hangdown'' geometry with a few-micrometer-wide microwire integrated into a coplanar waveguide (CPW) onto which a thin-film sample has been deposited \cite{Isaac2016apr}.
These protocols use voltages applied to the sample or cantilever to induce small electrostatic shifts in the cantilever's mechanical resonance frequency and, in parallel, use an optical fiber attached to the scanned-probe head to observe reflectance signals from patterned metallic features on the fixed sample stage.
We demonstrate aligning a high-compliance cantilever to the waveguide's microwire with micrometer precision in high vacuum at 295, 77, and 4.2 kelvin; the protocols work well even when the waveguide lies buried below a thin, anti-static metallic overlayer.
The protocols elaborated here recently allowed our team to observe --- for the first time in a single scanned-probe experiment --- electron-spin resonance, nuclear magnetic resonance, and  the creation of microwave-induced nuclear hyperpolarization \cite{Isaac2016apr}.

While the protocols described here were developed specifically for a magnetic resonance force microscope, the methods are potentially applicable to a wide range of high-sensitivity, high precision scanned-probe microscope experiments --- particularly those involving microcantilevers or nanowires to study precious samples, patterned samples, or metal coated devices \cite{Degen2009feb,Vinante2011dec,Longenecker2012nov,Nichol2012feb,Nichol2013sep,Grinolds2014mar,Tao2016sep,Rose2017jul}.

\section{Microscope description}

\begin{figure*}
    \includegraphics[width=6.25in]{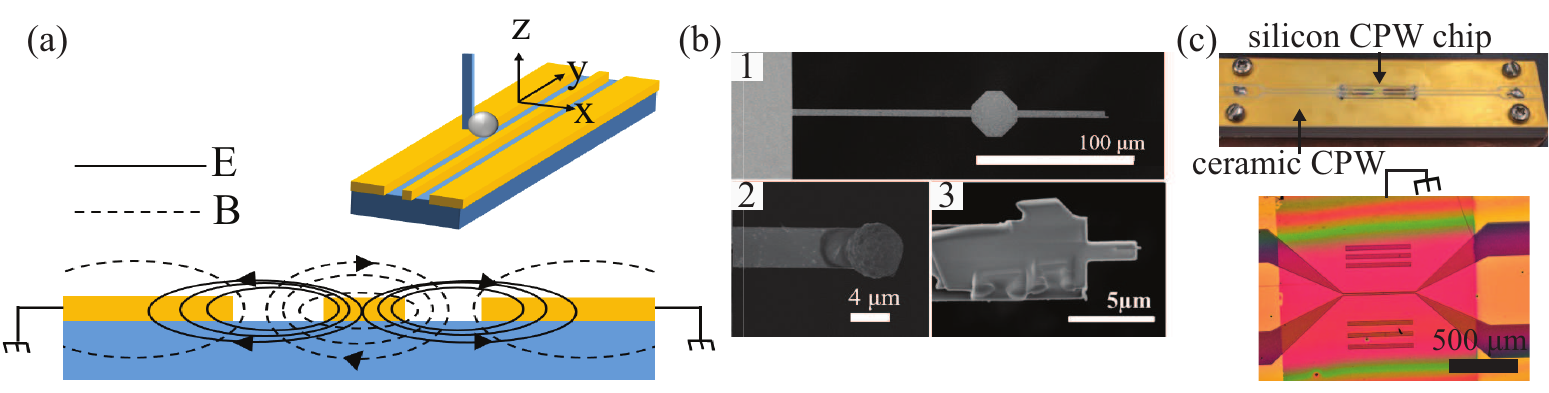}
    \caption{Experimental schematic. (a) Cross-sectional sketch of a coplanar waveguide showing the electric field lines (solid) and resulting magnetic field lines (dashed) expected when an oscillating voltage is applied to the centerline. The cantilever oscillates in the $x$-direction as shown. (b) Scanning electron microscope images of representative high-compliance silicon microcantilevers used including (1) a 200 $\micro\meter$ long cantilever with no magnetic tip \cite{Jenkins2004may}, (2) a silicon cantilever with a handglued, micrometer-scale nickel magnet tip, and (3) a silicon cantilever with a serially attached nanometer-scale cobalt magnet chip \cite{Longenecker2012nov}. (c) {\it {upper:}} Photograph of a coplanar waveguide (CPW) device showing the silicon CPW chip wirebonded into a ceramic CPW board. {\it{lower:}} Optical image of a coplanar waveguide with a 12 $\nano\meter$ thick gold coating over the center. This gold top contact was connected to the ground plane of the CPW via three aluminum wire bonds. }
    \label{fig:experimental_setup}
\end{figure*}

The following experiments were carried out on a custom-built magnetic resonance force microscope described in Ref.~\citenum{Isaac2016apr}.
The experimental geometry is sketched in Fig.~\ref{fig:experimental_setup}(a).
Briefly, we employed a custom-fabricated, high-compliance, silicon microcantilever \cite{Jenkins2004may} having either a micrometer-scale nickel magnet tip \cite{Garner2004jun} or a submicrometer-scale cobalt magnet tip \cite{Longenecker2011may,Longenecker2012nov} as a force detector (Fig.~\ref{fig:experimental_setup}(b)).
Cantilever motion was observed with a fiber-optic interferometer (wavelength $\lambda = \; 1310 \; \nano\meter$; estimated power $P = 2.5 \; \micro\watt$).
The interferometer output was fed into a controlled-gain phase-shifting circuit  whose output was used to drive a piezo element to which the cantilever was mounted.
The cantilever was thus driven into self-oscillation using this positive-feedback network \cite{Albrecht1991jan} and its frequency was measured using a commercial frequency counter (Stanford Research Systems SR620; Fig.~\ref{fig:freq_shift_sketch},~\ref{fig:freq_shift_align},~\ref{fig:alignment_under_metal}, and ~\ref{fig:optical_align}) or a software frequency demodulator  (Fig.~\ref{fig:alignment_under_metal}(c)) \cite{Dwyer2015jan}.

To perform magnetic resonance experiments, a magnetic field was applied along the $z$-direction with strength ranging from 0.6 to 1.4 $\tesla$ for electron spin resonance and up to 6 $\tesla$ for nuclear magnetic resonance. A coplanar waveguide (CPW) was used to manipulate spins by delivering a transverse oscillating magnetic field at frequencies ranging from $1 \; \hertz$ to $40\;\giga\hertz$.
The high field intensity was achieved by tapering the centerline of the waveguide to a narrow microwire.
The tapered coplanar waveguide enabled the delivery of a radiofrequency field with strength $B_1 \geq 2.5\;\milli\tesla$ (at frequencies between 25 and 250 $\mega\hertz$) and a microwave field of intensity $B_1 \geq 5\;\micro\tesla$ (at frequencies between 20 and 40 $\giga\hertz$) within the modest $200\;\milli\watt$ power budget of a liquid-helium cooled experiment.
The coplanar waveguide was fabricated by depositing 0.2 $\micro\meter$ copper capped with 0.03 $\micro\meter$ gold on a $10\;\milli\meter\times\;2\;\milli\meter$ high-resistivity silicon ($>$10,000$ \; \ohm \; \centi\meter$; $\epsilon_n = 11$ nominal) chip coupled via multiple {$38\;\micro\meter$ wide} aluminum wire bonds to a commercially fabricated ceramic (Rogers Corporation: TMM10i, $\epsilon_R$ = 9.8) CPW substrate equipped with SMA connections (Fig.~\ref{fig:experimental_setup}(c)).
The device resistance and performance was found to be dominated by the silicon chip itself and was largely unaffected by the composition, length, or number of wire bonds.
The silicon portion of the CPW was tapered from a $480\;\micro\meter$ wide copper/gold centerline with a $230\;\micro\meter$ silicon gap over a distance of $450\;\micro\meter$ to the microwire portion having a $10\;\micro\meter$ wide wire with a $6\;\micro\meter$ gap or a $5\;\micro\meter$ wide wire with a  $3\;\micro\meter$ gap and a length ranging from $20\;\micro\meter$ to $500\;\micro\meter$ (Fig.~\ref{fig:experimental_setup}(c)) \cite{Isaac2016apr}.
In all experiments shown, a $\sim \! 200\;\nano\meter$ thick nitroxide-doped polystyrene film sample was spun cast over the entire silicon coplanar waveguide.

To help mitigate surface frequency noise \cite{Kuehn2006apr} and decrease sample charging at cryogenic temperatures, some experiments employed a $12 \; \nano\meter$ thick gold coating deposited via electron beam evaporation over the polymer film in the tapered region of the coplanar waveguide (Fig.~\ref{fig:experimental_setup}(c)).
The metal thickness was estimated using a crystal monitor on the electron beam evaporator (CVC SC4500) and verified using atomic force microscopy (Veeco Dimension 3100).
A measured surface roughness of 1.8 nm rms --- equal to that of the underlying polymer film --- was determined by depositing the gold film through a grid composed of 35 $\micro\meter$ wide wire and 90 $\micro\meter$ diameter holes.
This mesh pattern enabled AFM scans across the gold/polymer edges in a 20 $\micro\meter$ scan range while reducing sample charging in the polymer regions.
A thickness of $12 \; \nano\meter$ was chosen for the film as it seemed thick enough to be continuous, yet thin enough to maintain a working distance of $\leq 50 \; \nano\meter$ between a nanomagnet tipped cantilever and the underlying sample.
The resulting metal layer was semi-transparent to visible light and had a $\sim \! 120 \; \Omega$ resistance when measured through 3 to 4 aluminum wire bonds connected to the CPW ground plane.
Additionally, the gold coating was successful in mitigating surface-induced cantilever frequency fluctuations with a carefully set cantilever-tip-bias voltage \cite{Stipe2001aug,Kuehn2006apr, Kuehn2008feb}.
Taken together, these observations imply the presence of a continuous metal coating over the polymer sample.

Magnetic resonance force microscope experiments were carried out at a temperature of $4.2\;\kelvin$ and a pressure of $5 \times \power{10}{-6} \; \milli\bbar$ \cite{Isaac2016apr}.
For these experiments the cantilever was oriented in the ``hangdown'' geometry (Fig.~\ref{fig:experimental_setup}), positioned over the microwire of the CPW, and brought within 10's of nanometers of the surface.
Three dimensional motion in our magnetic resonance force microscope was achieved using custom built Pan-style walkers for coarse motion and a piezo tube actuator for fine positioning \cite{Pan1993sep,Drevniok2012mar}.
At 4.2 $\kelvin$ the piezo tube actuator has a scan range of $\sim 1 \; \micro\meter$ in $z$ and $\sim 3 \; \micro\meter$ in the $x$ and $y$-directions.
The tip-sample separation was measured by applying a bias to the piezo tube actuator to gently touch the cantilever to the surface, and then removing the bias voltage.
We note that separate fiber optic interferometers were used to observe the $x$ and $y$ motion of the walkers simultaneously.
These two interferometers allowed us to record and correct for any undesired orthogonal-stage drift when scanning the cantilever stage in the lateral direction.
While the coarse positioning system allowed for several millimeters of motion in the $x$, $y$, and $z$-directions, this freedom in the system caused difficulties in the alignment upon cooling to cryogenic temperatures.
Large, unpredictable drifting due to thermal contraction allowed the cantilever to shift by hundreds of micrometers in an arbitrary direction between the setup at ambient conditions and operation at $4.2\;\kelvin$

\section{Alignment methods and results}

Below we describe two sets of methods for aligning a high-compliance cantilever to a 5 to 10 $\micro\meter$ wide wire integrated into a coplanar waveguide in a magnetic resonance force microscope: (1) methods in which cantilever frequency shifts are used to map the electrostatic forces associated with the underlying coplanar waveguide substrate and (2) methods in which a single optical fiber is used to obtain linescan images of the sample surface. Integrating various combinations of these techniques has proven useful depending on the sample of interest. To date, the electrostatic methods have been used when studying polymer thin-films both with and without a metal ground plane coating on the sample surface. The optical methods, while extremely useful in providing directional guidance and reducing the time required to approach the cantilever to the sample surface, have only been demonstrated on thin-film polymer samples in the absence of a metal coating. However, due to the transparency of the metal coating, we anticipate the optical methods described below would remain viable with a $\sim 10 \; \nano\meter$ metal surface coating on the sample.

\paragraph*{Cantilever frequency shift methods ---}

\begin{figure*}
    \includegraphics[width=6.25in]{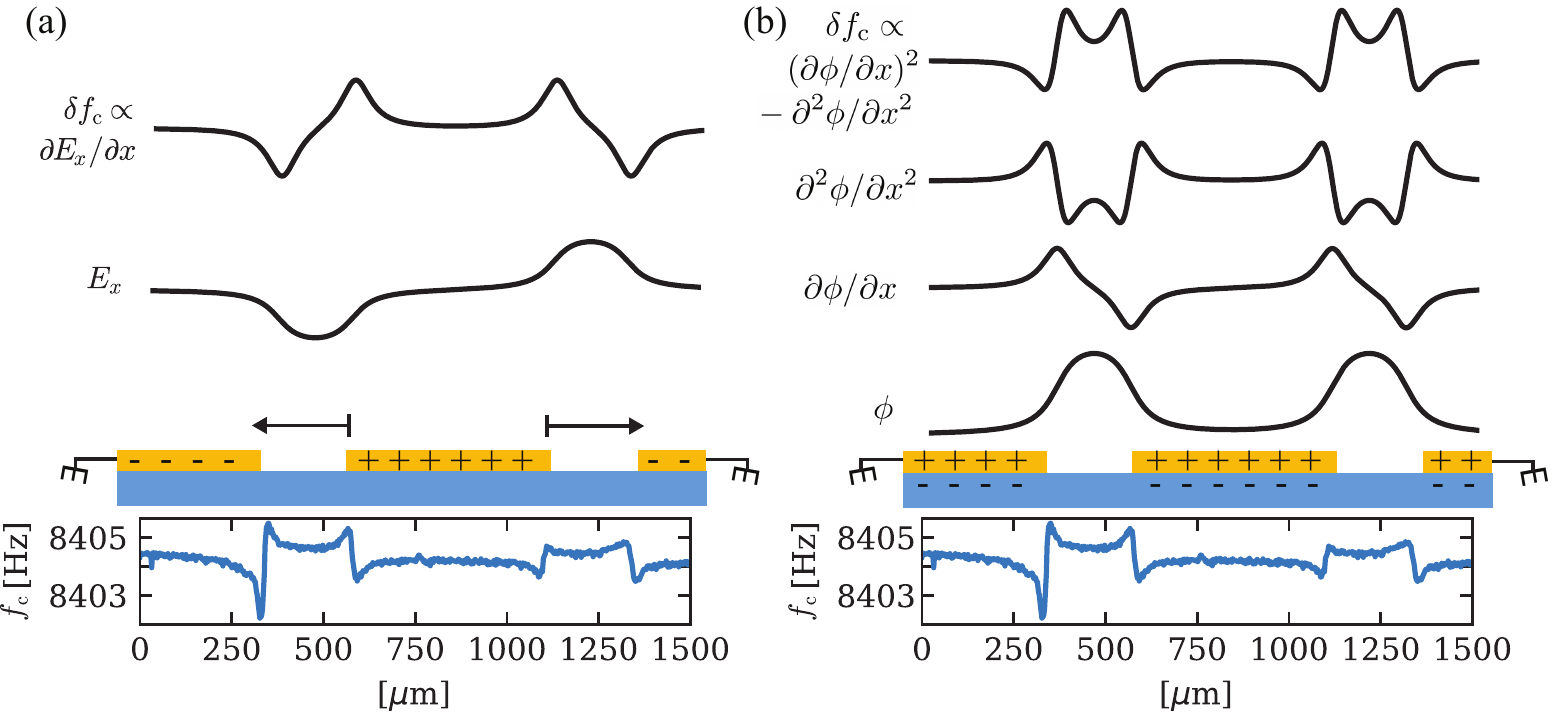}
    \caption{Proposed mechanisms for the cantilever frequency shifts observed while aligning a high-compliance cantilever, operated in the ``hangdown'' geometry, to a micrometer scale coplanar waveguide. (a) {\it{upper:}} A positive voltage applied to the CPW microwire generates an electric field $E_x$ as shown in the Fig.~\ref{fig:experimental_setup}(a). As the cantilever is scanned laterally across the surface, the electric field gradient $\partial E_x / \partial x$ interacts with charges $q$ on the cantilever tip to generate a force-gradient $\partial F_x/ \partial x$ which shifts the cantilever resonance frequency $f_{\mathrm{c}}$. (b) {\it{upper}}: A positive voltage applied to the CPW centerline builds up image charges in the underlying silicon substrate. These charges result in a contact potential difference $\phi$ between the conducting metal layer and the insulating silicon gaps of the CPW.  As the cantilever is scanned laterally across the surface, charges $q$ on the tip of the cantilever interact with a combination of the first and second derivatives of $\phi$ to generate the observed $\delta f_{\mathrm{c}}$. {\it{lower:}} A representative lateral scan of the cantilever across the untapered region of the CPW at 295 $\kelvin$. Experimental parameters: 200 $\mu$m long silicon cantilever with no magnet tip, cantilever spring constant $k_{\mathrm{c}} \sim \! 1.0 \; \milli\newton\;\meter^{-1}$, cantilever quality factor $Q \sim\! 15,000$; tip-sample separation $h\sim \! 3$ to 4 $\micro\meter$, cantilever bias voltage $V_{\mathrm{tip}} = -10 \; \volt$, coplanar waveguide centerline bias $V_{\mathrm{CPW}} = +10 \; \volt$.}
    \label{fig:freq_shift_sketch}
\end{figure*}

Figure~\ref{fig:experimental_setup}(a) shows a cross-sectional sketch of the coplanar waveguide and the cantilever; the cantilever oscillates in a direction perpendicular to the long axis of the CPW centerline.
To align the cantilever to the microwire section of the coplanar waveguide, we applied a bias voltage to the cantilever tip and/or the coplanar waveguide, self-oscillated the cantilever via positive feedback \cite{Albrecht1991jan}, and recorded the cantilever frequency with a hardware frequency counter as the cantilever was scanned laterally across the sample surface.
Figure~\ref{fig:freq_shift_sketch} (lower) shows a representative plot of the observed changes in cantilever resonance frequency as the cantilever was scanned in the $x$-direction across the wide, untapered region of the coplanar waveguide.
Characteristic shifts in the mechanical resonance frequency of the cantilever were observed along the gold/silicon interfaces.
It is of interest to note that these characteristic shifts in $f_{\mathrm{c}}$ were detectable at the gold/silicon interfaces with a bare silicon cantilever, and silicon cantilevers with both a micrometer scale nickel magnet tip and a nanometer scale cobalt magnet tip.
We also found that it was possible to detect small frequency shifts without applying a bias to the cantilever or coplanar waveguide.
These shifts were present at cantilever operating distances of several micrometers from the sample surface and were fairly insensitive to drift in the tip-sample separation or changes in sample topography.
Following these observations, we developed two hypotheses to explain the observed changes in cantilever frequency: (1) cantilever-tip charges interacting with the electric field emanating from the coplanar waveguide (Fig.~\ref{fig:freq_shift_sketch}(a)) and (2) cantilever-tip charges interacting with contact potential differences in the conducting metal and insulating silicon features of the coplanar waveguide (Fig.~\ref{fig:freq_shift_sketch}(b)).

For a high-compliance, beam cantilever oscillating parallel to a surface in the $x$-direction, Stowe {\it et al.} have derived a series of equations describing the effect of near-surface forces on the cantilever spring constant \cite{Stowe1997jul,Stowe2000}.
Assuming the surface force potential is conservative and the cantilever is oriented perpendicular to the surface at zero displacement, the effective spring constant $k_e$ is
\begin{equation}
k_e = k_{\mathrm{c}} - \dfrac{\partial F_x}{\partial x} - \dfrac{F_z}{\ell}
    \label{Eq:k_eff}
\end{equation}
where $k_{\mathrm{c}}$ is the intrinsic spring constant of the cantilever, $F_x$ is the force acting on the cantilever in the direction of oscillation ($x$), and $F_z$ is the force acting along the length $\ell$ of the cantilever.
\footnote{We note that there is a sign error in this equation as given in Ref.~\citenum{Stowe1997jul} that was corrected in Ref.~\citenum{Stowe2000}. The signs given in Eq.~\ref{Eq:k_eff} agree with the corrected version of this equation.}

Hypothesis 1 suggests that the observed changes in cantilever frequency resulted from cantilever-tip charges interacting with the lateral electric field associated with the coplanar waveguide as shown in Fig.~\ref{fig:experimental_setup}.
In this case, we expect the $z$-force term in Eq.~\ref{Eq:k_eff} to be negligible when $F_x$ and $F_z$ are comparable in magnitude and $h \ll \ell$, which is the case here, leaving $\Delta k = k_e - k_{\mathrm{c}} = \partial F_x / \partial x$. The force $F_x$ would be the result of a charge $q$ on the cantilever interacting with the $E$-field emanating from the coplanar waveguide, $F_x = q E_x$. Substituting this force into Eq.~\ref{Eq:k_eff}, we see that $\Delta k$, and therefore the cantilever frequency, resulting from interactions with a lateral electric field is dependent on the electric field gradient as
\begin{equation}
\Delta k_{E} = q \dfrac{\partial E_x}{\partial x}.
    \label{Eq:k_E}
\end{equation}
The frequency shift expected from this interaction is sketched in Fig.~\ref{fig:freq_shift_sketch}(a). This proposed mechanism is inconsistent with the observed frequency shift versus $x$ data.

Hypothesis 2 proposes a cantilever frequency shift resulting from surface potential differences between the insulating silicon and conducting metal layers of the coplanar waveguide. The surface potential energy $U$ contributing to the frequency shift in this mechanism, assuming a constant tip charge, is given by
\begin{equation}
U = -\frac{1}{2} C (V - \phi)^2
\label{Eq:surface_potential}
\end{equation}
where $C$ is the capacitance between the sample surface and the cantilever tip, $V$ is a voltage applied to either the cantilever or the sample, and $\phi$ is the potential difference between the two components of the system \cite{Silveira2007,Hoepker2011oct}. The resulting force on the cantilever $F_x = -\partial U / \partial x$ is
\begin{equation}
F_x = \dfrac{1}{2} \dfrac{\partial C}{\partial x} (V - \phi)^2 - C(V - \phi) \dfrac{\partial \phi}{\partial x}.
\label{Eq:Fx_phi}
\end{equation}
Again, we substitute this into our equation for $\Delta k$ and obtain
\begin{multline}
\Delta k_{\phi} = -2(V - \phi)\dfrac{\partial C}{\partial x} \dfrac{\partial \phi}{\partial x}
+ C \left(\dfrac{\partial \phi}{\partial x} \right)^2 \\
+ \dfrac{1}{2} (V - \phi)^2 \dfrac{\partial^2 C}{\partial x^2}
 - C (V - \phi) \dfrac{\partial ^2 \phi}{\partial x^2}.
\label{Eq:k_phi}
\end{multline}

For our operating distances of several micrometers, the contributions from $\partial C / \partial x$ and $\partial^2 C / \partial x^2$ over our fairly homogeneous sample substrate are assumed to be negligible. The resulting $\Delta k_{\phi}$ is therefore approximately
\begin{equation}
\Delta k_{\phi} \simeq C \left(\dfrac{\partial \phi}{\partial x} \right)^2 -  C (V - \phi) \dfrac{\partial ^2 \phi}{\partial x^2}.
\label{Eq:k_phi2}
\end{equation}
In Fig.~\ref{fig:freq_shift_sketch}(b), we sketch the expected cantilever frequency shift for $f_{\mathrm{c}} \propto (\partial \phi / \partial x)^2 - \partial ^2\phi / \partial x^2$ (i.e. ($V-\phi) \sim 1 \; \volt$) and find good agreement with the observed shape of the cantilever frequency transients.
While assuming a ($V-\phi$) value of 1 $\volt$ is reasonable for these experiments, changing this difference serves to alter the symmetry of the cantilever frequency shifts.
Increasing the value of ($V-\phi$) results in the cantilever frequency shifts becoming more symmetric while decreasing the magnitude of ($V-\phi$) results in a larger asymmetry.
Additionally, we note that if the sign of the bias voltage applied to the centerline or the cantilever was changed, the direction of cantilever frequency shifts was also inverted as expected from the second term in Eq.~\ref{Eq:k_phi2}.

\begin{figure}
    \includegraphics[width=3.37in]{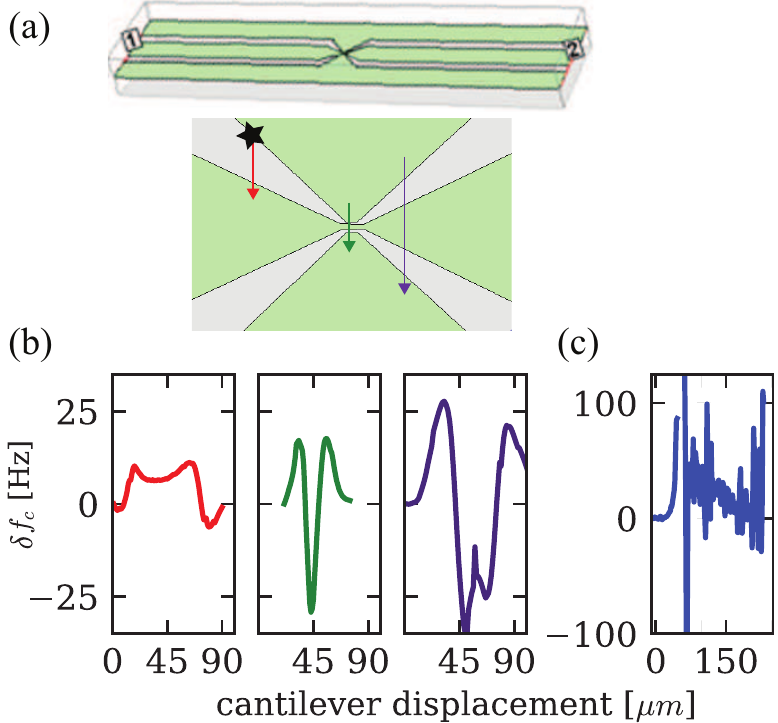}
    \caption{Detecting cantilever frequency shifts for cantilever alignment. (a) Top-down sketch of a coplanar waveguide. The input and output ports of the microwire are labeled as (1) and (2), respectively. Arrows indicate the location and direction of the scanning cantilever. (b) Observed cantilever frequency shift $\delta f_{\mathrm{c}}$ versus position, starting from the starred position in (a) and proceeding to the right. The first scan (red) shows two silicon/gold edges with the center of the scan over silicon. As the cantilever scans across the microwire region, the lineshape observed in Fig.~\ref{fig:freq_shift_sketch} becomes compressed as seen in the center (green) plot. As the cantilever moves back to the tapered region of the coplanar waveguide, the uncompressed lineshape is once again visible with the lower-frequency center region indicating the gold centerline (purple). Experimental parameters: 400 $\micro\meter$ long silicon cantilever with a $d \sim$ 4 $\micro\meter$ nickel magnet tip, $f_{\mathrm{c}} \sim \! 1460 \; \hertz$, $k_{\mathrm{c}}$ = 0.2 $\milli\newton\;\meter^{-1}$, $Q\sim \! 5,000$; temperature $T \; \sim \; 295$ kelvin, $h \; \sim \; 4 \; \micro\meter$, no metal sample coating, scanning near tapered region of a 10 $\micro\meter$ wide microwire CPW, $V_{\mathrm{tip}} = -4 \; \volt$, $V_{\mathrm{CPW}} = 0 \; \volt$. (c) Representative scan at 4.2 kelvin showing spurious shifts in $f_{\mathrm{c}}$. Experimental parameters: 200 $\micro\meter$ long silicon cantilever with a $d \sim \! 4 \; \micro\meter$ nickel tip, $f_{\mathrm{c}} \sim \!3540 \; \hertz$, $k_{\mathrm{c}} = 0.67 \; \milli\newton \; \meter^{-1}$; $h \sim \! 4 \; \micro\meter$, no metal coating on sample, $V_{\mathrm{tip}} = -4 \; \volt$. }
    \label{fig:freq_shift_align}
\end{figure}

\begin{figure}
    \includegraphics[width=3.37in]{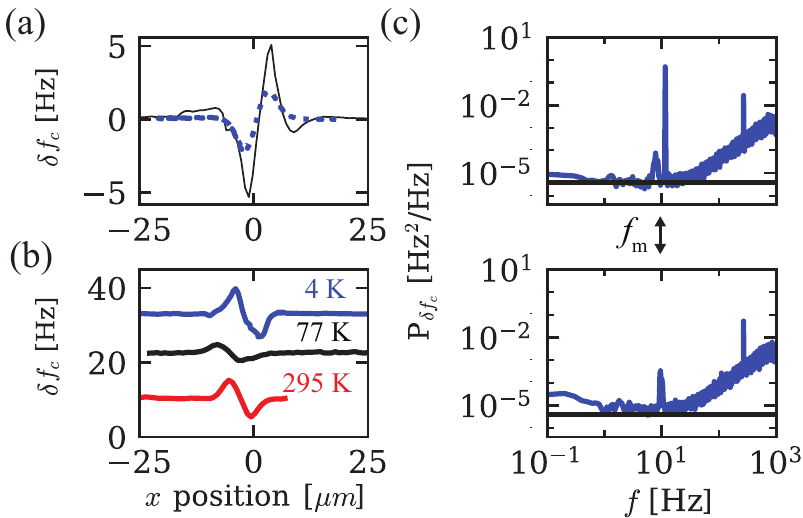}
    \caption{Aligning a nanomagnet tipped cantilever to a micrometer scale feature beneath a metal ground plane.
    (a) Cantilever frequency shift $\delta f_{\mathrm{c}}$ observed with a nanomagnet tipped cantilever over a 5 $\micro\meter$ wide microwire ($V_{\mathrm{CPW}} = -0.5 \; \volt$) integrated into a coplanar waveguide (CPW) at 295 $\kelvin$. The solid black line is the frequency shift observed in the absence of a gold metal coating on the sample while the blue, dotted line shows the resulting $\delta f_{\mathrm{c}}$ in the presence of a 12 $\nano\meter$ thick ground plane deposited over the sample surface. Both scans were obtained with $h \sim\! 5 \; \micro\meter$.
    (b) Observed $\delta f_{\mathrm{c}}$ with a metal coating on the sample and a +0.5 $\volt$ bias applied to the CPW microwire.
          The $y$-axes have been offset for clarity.
    (c) Cantilever frequency-shift power spectrum $P_{\delta f_{\mathrm{c}}}$ for a cantilever aligned over the CPW microwire (upper) and for a cantilever aligned over the flanking CPW ground plane (lower).
    An $f_m$ = 10 $\hertz$, +0.5 $\volt_{\mathrm{rms}}$ bias was applied to the CPW centerline in both instances.
	Experimental parameters: 200 $\micro\meter$ long silicon cantilever with $r$ = 67 $\nano\meter$ cobalt magnet tip, $f_{\mathrm{c}} \sim \! 6500 \; \hertz$, $k_{\mathrm{c}} = 2.0 \; \milli\newton \; \meter^{-1}$, $Q \sim \! 20,000$ ($295 \; \kelvin$) to $\sim \! 70,000$ ($4.2 \; \kelvin$); $h$ $\sim$ 3 to 4 $\micro\meter$; sample coated with a $12 \; \nano\meter$ thick gold layer; scans over 5 $\micro\meter$ wide microwire region of coplanar waveguide.}
    \label{fig:alignment_under_metal}
\end{figure}

This positioning method was used to successfully and reproducibly align the cantilever with respect to the waveguide at $295 \; \kelvin$ and $77 \; \kelvin$.
Fig.~\ref{fig:freq_shift_align} shows lateral scans over various regions of the coplanar waveguide showing how this method was used to map out the surface of the CPW. In Fig.~\ref{fig:alignment_under_metal}(a), we show a scan over a 5 $\micro\meter$ wide microwire.
The observed lineshape is reproducible but does not show all of the edge effects observed in Figure~\ref{fig:freq_shift_sketch} and Figure~\ref{fig:freq_shift_align}.
This lineshape is still consistent with the mechanism of hypothesis 2.
At large tip sample separations, the capacitance and tip-sample interactions are dominated by the 5 micrometer wide cantilever tip --- on the order of or larger than the sample features of the 5 $\micro\meter$ wide wire and 3 $\micro\meter$ wide silicon gaps --- leading to a `blurring' of the sample features.
At $4.2 \; \kelvin$ charge accumulation within the polymer thin film resulted in large, spurious cantilever frequency shifts that masked the desired frequency-shift signal (Fig.~\ref{fig:freq_shift_align}(c)).

In an attempt to mitigate this sample charging  observed at $4.2 \; \kelvin$, the sample was coated with a $12 \; \nano\meter$ thick gold coating.
We expected the gold ground plane to shield the cantilever from any electrostatic fields arising from the buried CPW microwire \cite{Fahy1988nov, Bailleul2013nov}; nevertheless, a characteristic microwire-related cantilever frequency shift was cleanly observed, over temperatures ranging from $295$ to $4 \; \kelvin$ (Fig.~\ref{fig:alignment_under_metal}(a, b)). As shown in Fig.~\ref{fig:alignment_under_metal}(a), the lineshape of the cantilever frequency shift changes slightly when the metal top contact is added, suggesting a slightly different mechanism.
For both experiments, with and without the metal coating, an operating distance of $h = 3$ to $5 \: \; \micro\meter$ between the cantilever tip and sample was used with a $+0.5 \; \volt$ bias applied to the centerline of the coplanar waveguide.
While there is a height difference between the $0.20 \; \micro\meter$ thick copper forming the CPW microwire/ground plane and the underlying silicon substrate, atomic force microscopy and contact profilometry (Alpha Step 500 Stylus Profilometer), indicate the polymer sample and gold top contact coated the waveguide conformally.
Additionally, for the $h$ used throughout, we would expect this $<5\%$ change in tip-sample separation to contribute less than 1 $\hertz$ to $\delta f_{\mathrm{c}}$.
These findings suggest that the observed 6 to 8 $\hertz$ shift in cantilever frequency (Fig.~\ref{fig:alignment_under_metal}) does not result solely from changes in $h$.
As in previous experiments without the ground plane between the sample and cantilever, the cantilever frequency shift does change sign with a change in the sign of the applied bias suggesting some electrostatic origins.
While the mechanism underlying the frequency shift observed over a buried feature is puzzling, it is nevertheless reproducible and a useful tool for finding the coplanar waveguide microwire.

To confirm the cantilever's alignment with the coplanar waveguide microwire and avoid misinterpreting spurious cantilever frequency shifts, we developed a modulated detection scheme.
Rather than applying a DC bias to the CPW, we applied an $f_{\mathrm{m}}$ = 10 $\hertz$ voltage and recorded the cantilever frequency-shift power spectrum $P_{\delta f_{\mathrm{c}}}$. As seen in Fig.~\ref{fig:alignment_under_metal}(c), a peak several orders of magnitude above the thermal noise floor was observed at the modulation frequency when the cantilever was near the CPW microwire. This modulation peak decreased in magnitude as the cantilever was scanned away from the CPW microwire and was absent at a lateral distance of 10 to 20 micrometers.
Again, we believe this modulation signal originates from cantilever tip charges interacting with an electric field from the coplanar waveguide rather than capacitive forces as it scales linearly with the magnitude of the voltage applied the CPW.

\paragraph*{Optical alignment methods}

\begin{figure}
    \includegraphics[width=3.37in]{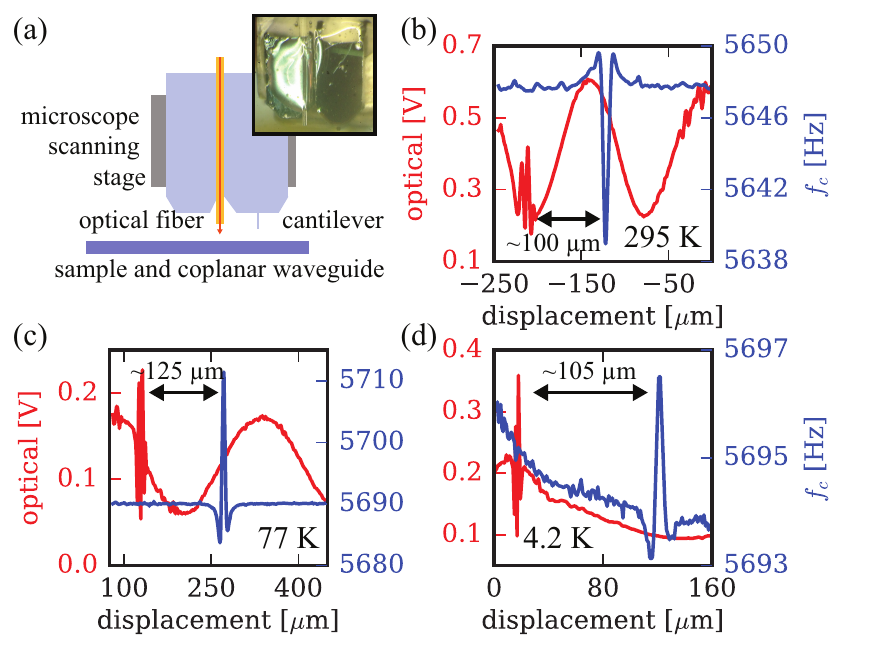}
    \caption{
    Aligning a high-compliance cantilever to a micrometer-scale coplanar waveguide using optical reflection and cantilever frequency shifts in tandem.
    (a) Sketch showing the placement of an optical fiber in proximity of the cantilever, pointed at the sample surface; (inset) a photograph of the experimental setup. Reflected optical signal (red, left) and cantilever frequency (blue, right) versus position at
    (b) 295~kelvin,
    (c) 77~kelvin, and
    (d) 4.2~kelvin.
    The optical signal corresponding to the CPW microwire has a distinct lineshape resulting from the gold/silicon/gold/silicon/gold pattern of the waveguide.
  	The sine wave observed in the background of the optical signal is the interferometer signal resulting from drift in the tip-sample separation over the lateral scan range.
	The cantilever frequency signal shows the same derivative lineshape seen in Fig.~\ref{fig:freq_shift_align}(b).
    Experimental parameters: 200 $\micro\meter$ long silicon cantilever with $r = 1.5 \; \micro\meter$ nickel tip, $k_{\mathrm{c}} \sim \! 3.0 \; \milli\newton \; \meter^{-1}$, $Q$ $\sim$ 22,000 at $295 \; \kelvin$ to 70,000 at $4.2 \; \kelvin$; $h$ $\sim\! 4$ to 6 $\micro\meter$; sample not coated with metal layer; scans over the narrow 10 $\micro\meter$ wide microwire region of CPW. At 295 $\kelvin$: $V_{\mathrm{tip}} = -4 \; \volt$, $V_{\mathrm{CPW}} = -1 \; \volt$; At 77 $\kelvin$: $V_{\mathrm{tip}} = 0 \; \volt$, $V_{\mathrm{CPW}} = -1 \; \volt$; At 4 $\kelvin$: $V_{\mathrm{tip}} = 0 \; \volt$, $V_{\mathrm{CPW}} = -0.25 \; \volt$.
	}
    \label{fig:optical_align}
\end{figure}

The cantilever-frequency-based methods described above have proven extremely useful in definitively aligning an attonewton sensitivity cantilever to the 5  to 10 $\micro\meter$ wide wire of our CPW device.
While these methods are reliable, they provide little insight into the cantilever's location over the CPW when the cantilever is far from the surface or positioned over the CPW's flanking ground plane.
To overcome these limitations and dramatically decrease the time required to position the cantilever both laterally and vertically, we developed a single fiber optical imaging protocol.
This simple setup, shown in Figure~\ref{fig:optical_align}(a), consists of an optical fiber and a photodetector.
The cleaved end of the optical fiber was affixed to a silicon chip for ease of positioning and placed on the microscope scanning stage adjacent to the cantilever, and pointed toward the sample surface.
The fiber was aligned such that the cantilever-sample separation and fiber-sample separation were approximately equivalent.
This setup did not require any lenses to focus or collect the laser light reflected off the surface, allowed for rapid approach to the surface and demonstrated micrometer resolution in distinguishing silicon and gold sample features.

Far from the surface, the magnitude of the optical signal was comparable to the amount of light reflected off the cleaved end of the fiber.
This surface-reflection signal increased rapidly, however, as the fiber came within a few micrometers of the surface --- allowing for a rapid approach of the cantilever to the surface without the risk of crashing the tip.
A final, slow approach to the surface could then be made over a much smaller (a couple $\micro\meter$) distance.
By watching the interferometer oscillations in the optical fiber, any stage drift in the vertical direction or sample offset angle could be measured.
A very distinct reflection signal was observed when the optical fiber was scanned laterally across the gold/silicon/gold pattern of the CPW as seen in Fig.~\ref{fig:optical_align}.
Using the methods described in the prior section to calibrate the distance between the optical signal and cantilever frequency shift signal at room temperature (Fig.~\ref{fig:optical_align}(b)), an offset vector was determined.
The cantilever was moved in the $x$/$y$-directions by an amount equal to this offset vector at cryogenic temperatures for rapid lateral alignment of the cantilever to the CPW.
This hybrid optical/electromechanical alignment protocol provided two advantages:
	(1) it increased the scan speed by a factor of two --- from 200 to 400 $\nano\meter / \second$  and
	(2) it reduced the risk of sample charging by eliminating the need for applying a bias voltage to the cantilever or CPW at $4.2 \; \kelvin$.
The increased scan rate came from eliminating the 2 to 3 second wait time required after each step of the Pan walker before reading the cantilever frequency. The abrupt motion of the coarse stepper induced cantilever ringing artifacts that made it difficult to accurately determine the cantilever resonance frequency immediately after moving. This motion-induced cantilever amplitude effect could, alternatively, be reduced by using amplitude feedback control.

\begin{figure}
    \includegraphics[width=3.37in]{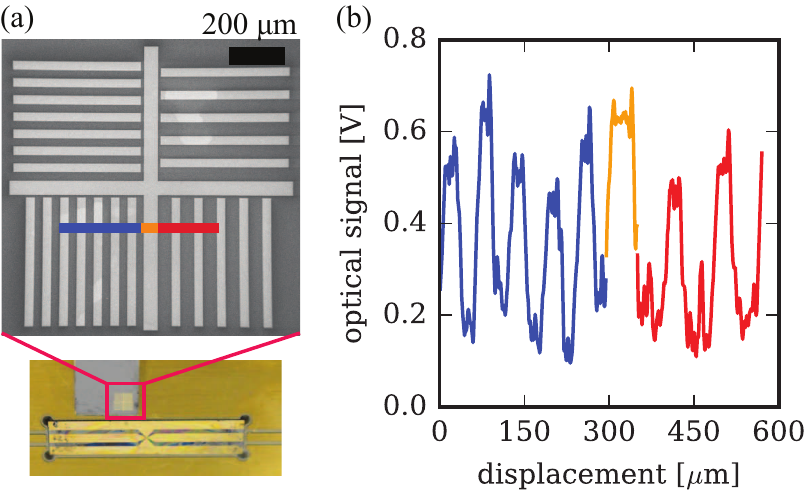}
    \caption{Using a microfabricated reticle.
    (a) Scanning electron micrograph and optical photograph of gold/silicon reticle.
Overlaid is a line indicating the regions of different dimensions.
    (b) Reflected optical signal versus lateral displacement of the optical fiber.
The spacing between and widths of the intense signal correspond to the various regions of the optical pattern shown in (a).
The left (blue) region has 30 $\micro\meter$ gold lines separated by 30 $\micro\meter$ silicon, the center (orange) gold line is 50 $\micro\meter$ wide, and the right (red) region has 30 $\micro\meter$ gold lines separated by 50 $\micro\meter$ of silicon.}
    \label{fig:grids_optical_ref}
\end{figure}

To further assist in the alignment, we developed a reticle to help determine the direction of motion required to bring the cantilever and waveguide into alignment.
As seen in Fig.~\ref{fig:grids_optical_ref}, we fabricated $1 \;\milli\meter \times 1 \; \milli\meter$ silicon chips patterned with $50 \; \nano\meter$ thick gold grid lines. The pattern consisted of $30$ $\micro\meter$ wide gold lines spaced by either $30$ or $50$ $\micro\meter$ of bare silicon.
The various quadrants were separated by a 50 $\micro\meter$ wide gold line.
The optical signal shown in Fig.~\ref{fig:grids_optical_ref}(b) demonstrates the resolution of this single optical fiber imaging.
The regions of the chip can be easily differentiated with the more intense reflection signal corresponding to regions of gold and the weaker signal corresponding to regions of silicon.
Again, by calibrating the position of the optical fiber and cantilever at room temperature, this grid system could direct the motion for alignment under operating conditions.
To eliminate the complication of aligning an external grid chip and an optical fiber, a similar grid system was implemented in the ground plane of the CPW as seen in Figure~\ref{fig:experimental_setup}(c).
These added features reduced the amount of ``signal-less'' area when scanning the optical fiber/cantilever and the reticle feature spacing provided guidance on which direction to move when aligning the cantilever to CPW.

\paragraph*{Mechanical stop ---}

\begin{figure}[b!]
\includegraphics[width=\columnwidth]{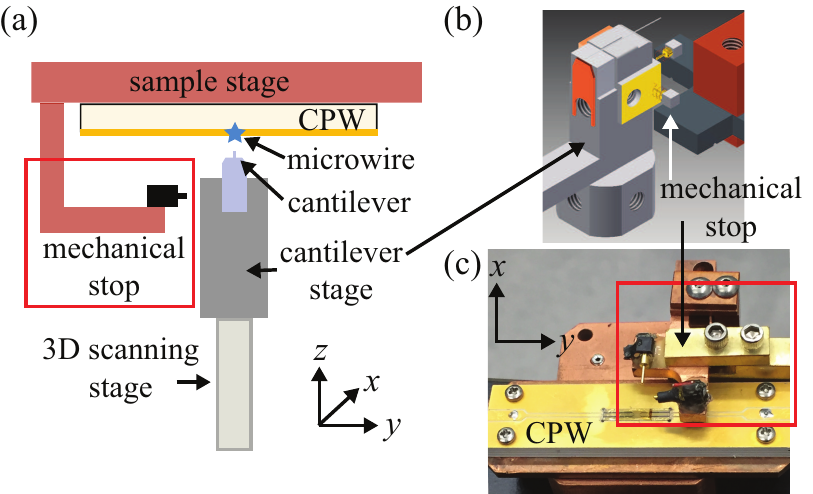}
\caption{Implementing a mechanical stop. (a) Side-view sketch of the microscope setup with the mechanical stop shown (boxed in red; not drawn to scale).(b) A 3-dimensional sketch of the cantilever stage and the mechanical stop showing the stops in both the $x$ and $y$ directions. (c) A photograph of the mechanical stop used in experiments (boxed in red) showing its position relative to the coplanar waveguide. The cantilever stage is omitted for clarity.
\label{Fig:probe_stops}}
\end{figure}

As a fail-safe, an L-shaped mechanical stop was affixed to the sample stage as shown in Fig.~\ref{Fig:probe_stops}.
This mechanical stop provided a corner as a point of reference in the $x$ and $y$ directions to guide motion and reduced the scan range of the Pan-style walker by several millimeters.
Under ambient conditions, the stop was positioned to come into contact with the cantilever stage when the cantilever was $\sim$100 to 200 $\micro\meter$ from alignment with the microwire.
Under vacuum, the cantilever was aligned to the microwire using the cantilever frequency shift methods described previously.
After the cantilever was aligned to the microwire, the stage was moved toward the stop.
It was determined that the cantilever stage was in contact with the mechanical stop when the voltage required to move the Pan-style walker increased significantly.
The distance between the microwire and the mechanical stop, in both the $x$ and $y$ directions, was recorded.
This distance remained unchanged at cryogenic temperatures as the mechanical stop and sample substrate were both firmly affixed to the same copper plate.
Therefore, if thermal drift of the cantilever stage caused misalignment upon cooling to $4.2\;\kelvin$, the cantilever stage could be quickly moved toward the stop until contact was made.
Once the stage was in contact with the stop, it was clear which direction, and how far, the cantilever stage needed to be moved to align the cantilever to the CPW microwire.
This mechanical barrier to motion allowed for rapid scanning of the sample stage and provided a safeguard if any of the previously described alignment methods failed upon cooldown.

\section{Conclusions}

Approaching a cantilever to a surface and aligning to features of interest is necessary in many scanned-probe experiments.
Additionally, having the ability to locate features buried beneath a metal coating may prove useful in studying electronic devices in scanning probe microscopy experiments.
Protocols to align a cantilever and sample rapidly and reproducibly are critical for experimental throughput and accuracy.
Here we introduced two classes of methods for reliable cryogenic alignment with micrometer precision:
	(1) cantilever frequency shift protocols applicable to experiments involving a high-compliance cantilever operating in the ``hangdown'' geometry and
	(2) optical-fiber-based alignment protocols applicable to essentially any scanning probe microscopy experiment.
At low temperature, full optical imaging of the cantilever and waveguide is prohibitively complicated;  the use of a single optical fiber, in contrast, was found to be straightforward.
Upon introducing these techniques into our magnetic resonance force microscopy experiments, we have decreased the alignment time under operating conditions ($4.2 \; \kelvin$, $5 \times \power{10}{-6} \; \milli\bbar$) from several days to just a couple of hours.
Additionally, we have increased the reproducibility of our alignment to the coplanar waveguide microwire.
These dramatic improvements have allowed for higher throughput and new experiments such as the integration of electron spin resonance, nuclear magnetic resonance, and dynamic nuclear polarization into a single scanned-probe experiment for the first time \cite{Isaac2016apr}.

\begin{acknowledgments}

This work was funded by the Army Research Office (grant no.\ W911NF-12-1-0221) and the National Science Foundation through Cornell's GK12 Program (grant no.\ DGE-1045513). This work was performed in part at the Cornell NanoScale Facility, a member of the National Nanotechnology Coordinated Infrastructure (NNCI), which is supported by the National Science Foundation (grant no.\ ECCS-115420819). This work made use of the Cornell Center for Materials Research Facilities supported by the National Science Foundation (award no.\ DMR-1719875).

\end{acknowledgments}

%


\label{TheEnd}
\end{document}